\documentclass{iauc}
\usepackage{graphicx}
\usepackage{natbib}
\input{epsf.sty}
\pubyear{2005}
\volume{199}
\pagerange{1--}
\jname{Probing Galaxies through Quasar Absorption Lines}
\editors{P. R. Williams, C. Shu, and B. M\'{e}nard, eds.}
\newcommand{\solar}{\ifmmode_{\mathord\odot}\else$_{\mathord\odot}$\fi} 
\newcommand{\kms}{km\thinspace s$^{-1}$}     
\newcommand{\arcs}{\ifmmode {'' }\else $'' $\fi}  
\newcommand{\arcm}{\ifmmode {' }\else $' $\fi}    
\newcommand{\mstar}{\ifmmode {M_{HI_\ast}}\else $M_{HI_\ast}$\fi}
\newcommand{\msolar}{M$_\odot$}

\title[HI Properties and Environment of Lyman-$\alpha$ Absorbers] 
{The HI Properties and Environment of Lyman-$\alpha$ Absorbers}

\author[Rosenberg]   
{Jessica L. Rosenberg$^{1,2}$}

\affiliation{$^1$Harvard-Smithsonian Center for Astrophysics 
\break 60 Garden Street MS 65 \break Cambridge, MA 02138, 
USA \break email: jlrosenberg@cfa.harvard.edu\\[\affilskip]
$^2$NSF Astronomy and Astrophysics Postdoctoral Fellow}

\begin{document}

\maketitle

\begin{abstract}

We present results from two projects in which we have used the HI 21cm emission 
line as a tracer of gas-rich galaxy populations in the vicinity of
Lyman-$\alpha$ absorbers. In the first case, we examine the HI environment of SBS
1543+593, the nearest damped Lyman-$\alpha$ absorber. We use a VLA map of the
region around this LSB galaxy which itself shows an extended
HI disk to identify 3 gas rich neighbors within 185 kpc. While it is not clear
whether we should expect local damped Lyman-$\alpha$ systems to reside in such
gas-rich regions, we would expect this kind of environment to be more prevalent
at higher redshifts where less of the gas is in the dense inner regions of
galaxies or has been consumed by star formation. This local galaxy is the only
system in which we can study the gaseous environment in this kind of detail. In 
the second case, we examine the HI environment surrounding 16 Lyman-$\alpha$ forest 
absorbers along 4 QSO sightlines. We do not detect any gas-rich galaxies at the 
absorber positions indicating that, at least down to our sensitivity limits,
these absorption lines do not seem to be associated with galaxy halos. For half
of the Lyman-$\alpha$ absorption systems there is a galaxy within 500 kpc, but 
for the other half there is not. In two cases there is no galaxy within 2 Mpc of
the Lyman-$\alpha$ absorption systems indicating that absorbers do, in some
cases, reside in voids. 


\end{abstract}

\firstsection 
\section{Introduction}

Lyman-$\alpha$ absorbers of all varieties are overdensities in the gas
distribution in the universe. The highest overdensities are the damped
absorbers which trace the gas in galaxies and protogalaxies through 
cosmic time. At the low column density end (N$_{HI} < 10^{17}$ cm$^{-2}$), 
the Lyman-$\alpha$ forest traces the smaller overdensities in the gas.

Our understanding of the formation and evolution of galaxies is closely tied to 
our understanding of their environment. The HI environment of galaxies is 
particularly important as it represents the primordial material that can fall
into galaxies and the processed gas that has been ejected. In addition, HI
provides an important complement to optical observations of galaxies since many 
of the systems detected at 21 cm are missed in optical surveys (particularly the
spectroscopic surveys that allow for a redshift determination) because they are 
low surface brightness (LSB) systems.  

I discuss two very different projects in which we use 21 cm HI observations to
survey the environments of Lyman-$\alpha$ absorbers. In the first case we look
at the environment of the nearest damped Lyman-$\alpha$ (DLA) system, SBS 1543+593,
and in the second case we look at the environment surrounding 16 low redshift
Lyman-$\alpha$ forest absorbers. 

\section{The HI Environment of SBS 1543+593}

As the closest DLA system outside of the local group, 
SBS 1543+593 gives us a rare opportunity to study one of these systems in great 
detail. We use HI observations made with the Very Large Array (VLA) in
C-array (E. Brinks, PI; D. Bowen and T. Tripp collaborators) to study the gas
distribution in and around SBS 1543+593.

We use standard AIPS data reduction techniques and create a map with a 15\arcs
$\times 14$\arcs\ CLEAN beam. We use these data to derive the HI parameters for
SBS 1543+593 and its neighbors and to make the HI contour map shown in Figure 1. We
measure an HI mass of 1.2$\times 10^9$ M\solar\ for SBS 1543+593, consistent
with the values measured by \citet{chengalur2002} and \citet{bowen2001b}. In
addition, the HI distribution shown in Figure 1 is consistent with the map of
\citet{chengalur2002} including the HI hole in the central region of the galaxy,
a dense HI ring, and HI spurs to the north and south of the galaxy's center. 
However, there are several interesting features of this galaxy when the optical
and HI maps are compared. Figure 1 shows that the HI disk is much 
larger than the optical extent. In fact, the optical extent of the galaxy 
corresponds with the dense ring seen in the 21 cm emission map. We identify a
spur of HI emission off the NNW end of the ring that corresponds to a faint 
optical extension while a spur of HI emission to the SSE of the ring also has
corresponding faint optical emission. 

\begin{figure}[ht]
\begin{center}
\epsfxsize=4.5in
\leavevmode
\epsffile{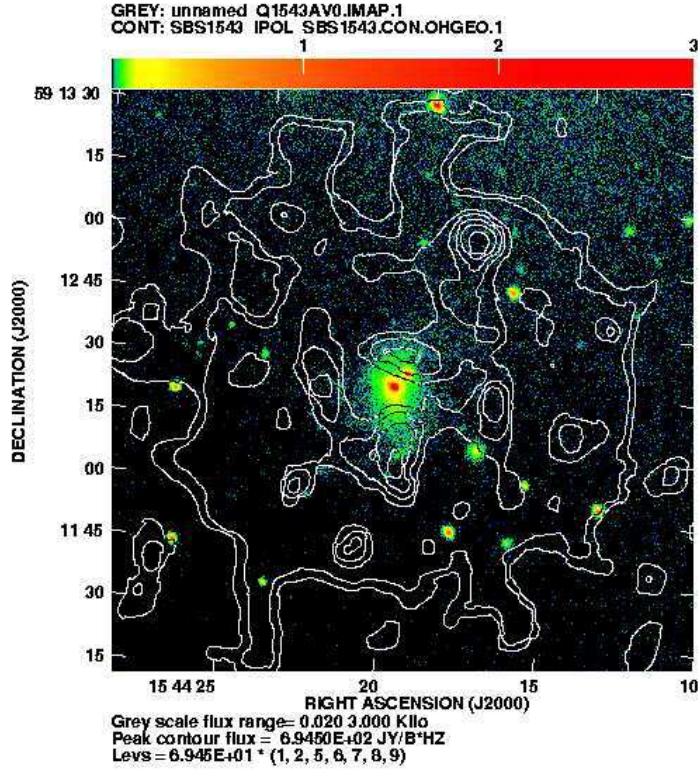}
\end{center}
\caption{An optical R-band image of SBS 1543+593 overlayed with HI contours
from the VLA map. The lowest level contour represents 3$\times 10^{20}$ 
cm$^{-2}$ column density. Note the large
extent of the HI disk well beyond the optical LSB galaxy. The
optical galaxy corresponds to the dense central ring of gas in this system.
Spurs off the north and south sides of the disk may be a result of tidal
interactions with the neighboring galaxies.}
\end{figure}

In surveying the region immediately surrounding SBS 1543+593, we identify 3 
gas-rich galaxies. Two of the galaxies were not previously cataloged while the
third is MCG+10-22-038 which, while previously known, did not have a previously
measured redshift. These galaxies reside 183 kpc (MCG+10-22-038),
161 kpc, and 123 kpc from SBS 1543+593 and have HI masses of 3.7$\times 10^8$ 
M\solar, 2.2$\times 10^8$ M\solar, and 6.1$\times 10^8$ M\solar\ respectively.

For the small detection volume of this study covered by the VLA
map, a very conservative estimate of the average galaxy density would predict
8.6$\times 10^{-3}$ galaxies down to log(M$_{HI}$/\msolar) = 8.07 in the field 
using the HI mass function from \citet{rosenberg2002}. Since this region is 
clearly not an unbiased position in the field since it was centered around a
known galaxy and galaxies tend to cluster, a higher than average galaxy
density should not be surprising. However, the detection of 3 galaxies in the
immediate vicinity of SBS 1543+593 does indicate a significant overdensity that
might not be expected near a LSB galaxy since they tend to be
less clustered than their higher surface brightness counterparts on large scales 
\citep{mo1994}. However, clustering around low surface brightness galaxies on 
scales less than 0.5 Mpc is highly variable with 20\% of systems having 3 or
more near neighbors \citep{bothun1993}. At least 2 of the three systems that we
detect near SBS 1543+593 probably would not have been included in the 
\citet{bothun1993} so it is not clear how often LSBs have neighbors like these
dwarf galaxies.

Figure 1 shows that SBS 1543+593 has an HI disk that is extended well beyond the
optical radius of the galaxy. The large extent of the HI disk is
surprising given the high density of the galaxy's environment. Nevertheless, the
HI distribution in SBS 1543+593 shows evidence for spurs in the outer part of
its disk and the HI distributions of the neighbors also show disturbed
morphologies possibly indicative of tidal disruption. 

Figure 2 shows the distribution of galaxies from the ZCAT catalog, which includes
data from the CfA survey as well as from several other galaxy surveys, 
(http://cfa-www.harvard.edu/~huchra/zcat/) in the region surrounding SBS 1543+593 
(small filled circles). The dashed line in the figure indicates the line of 
sight to the QSO HS 1543+5921 in which the damped Lyman-$\alpha$ absorption from 
SBS 1543+593 was identified. The points indicate 
that these new galaxies (large filled circles) are closer to SBS1543+593
(triangle) than any of the previously known galaxies. The presence of such
a large reservoir of primordial HI in the disk of SBS 1543+593 and in the
neighboring dwarf galaxies seems to indicate that this is a very young region.
One would expect the tidal fields that are distorting the HI distribution in all
of these galaxies to trigger substantial star formation but there is no evidence
for this optically.  

\begin{figure}[ht]
\begin{center}
\leavevmode
\epsfxsize=2.6in
\epsffile{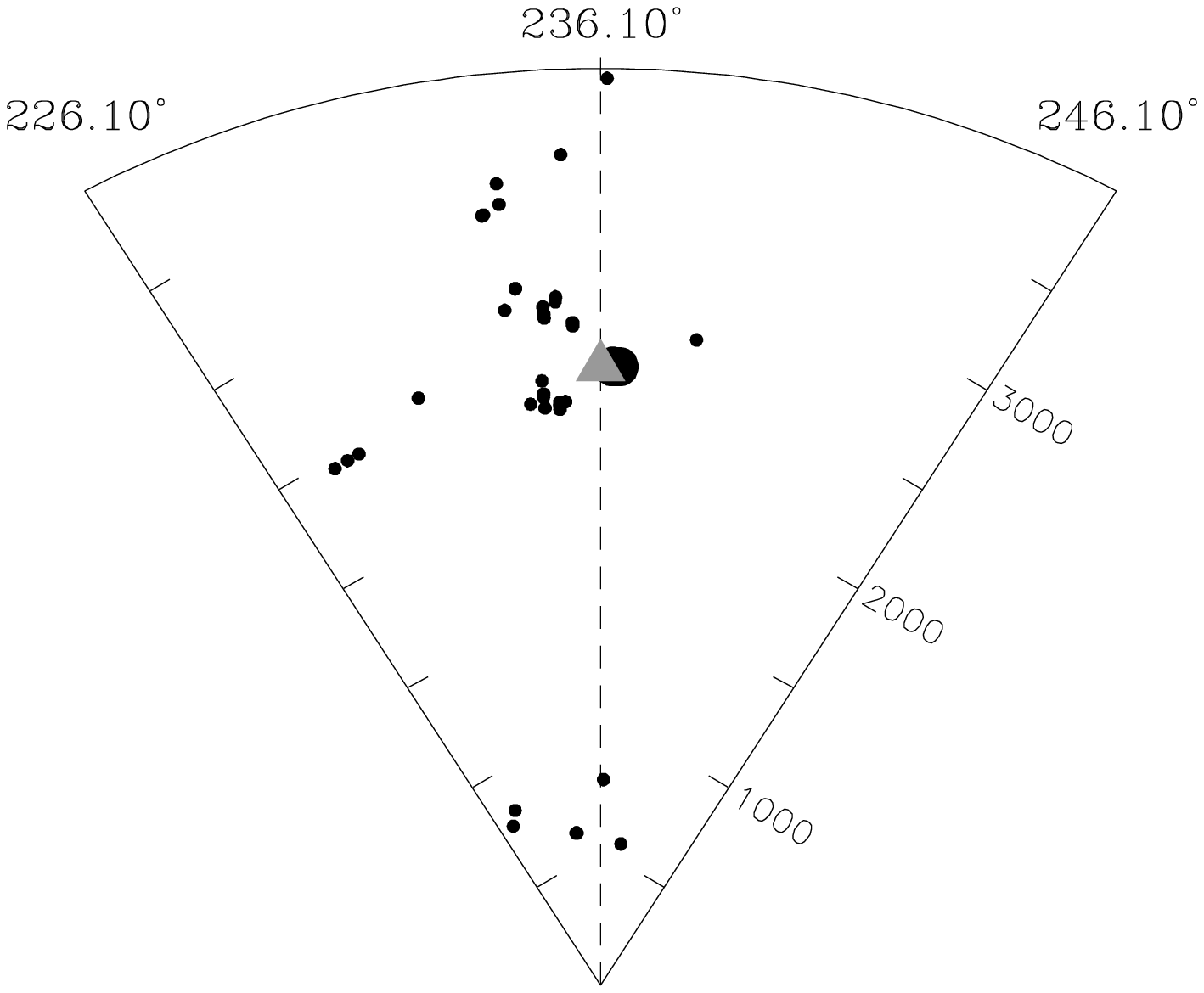}
\epsfxsize=2.6in
\epsffile{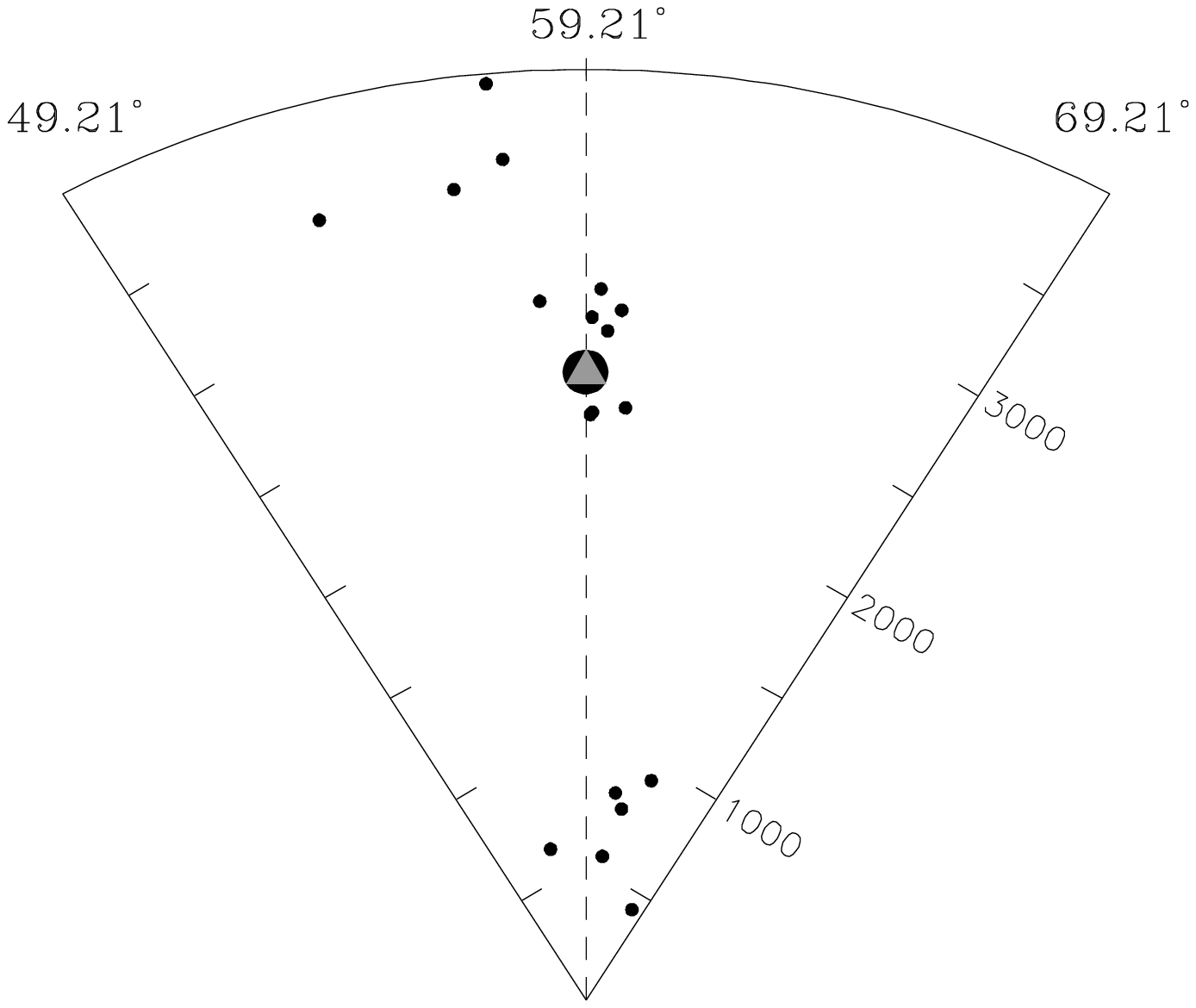}
\end{center}
\caption{The large scale distribution of galaxies around DLA SBS 1543+593 which is
shown as the gray triangle. The large filled circles represent the HI detections 
from the VLA maps, and
the smaller filled circles represent galaxies compiled in ZCAT 
(http://cfa-www.harvard.edu/~huchra/zcat/) which includes the CfA redshift survey
and other galaxy survey information.}
\end{figure}

The neighbors to SBS 1543+593 range in HI diameter at a column density
of 2$\times 10^{20}$ cm$^{-2}$ from 7 -- 15 kpc as
compared with the 27 kpc diameter of SBS 1543+593. Clearly the cross-sections of 
these galaxies are much less significant than that of the DLA. However, one
might expect these HI-rich dwarfs to be more common at higher redshift where the
multiple DLAs are observed. While SBS 1543+593 is a single system that can not
be used to predict the environments of other DLAs, we must factor into our
interpretation that higher
redshift systems may reside in equally dense or denser HI environments where we
would not be able to detect the gas-rich neighbors.

\section{The HI Environment of Lyman-$\alpha$ Forest Absorbers}

Lyman-$\alpha$ forest absorbers are the lowest column density systems and
therefore provide a tracer of the smallest overdensities in the HI distribution.
These systems have given rise to tremendous debate about their nature and their
association with galaxies. Some groups have suggested that these absorbers are 
associated with LSB galaxies \citep{impey1997} while others argue that they trace 
the extended halos of high surface brightness galaxies \citep{Lin2000, 
Lanzetta1995} or that they trace filaments of primordial material that
are correlated with galaxies because they are overdensities in the same large
scale structures \citep{dave1999, stocke2000}.

We have used HI 21cm observations from Arecibo, Parkes, and the Australia
Telescope Compact Array to search for gas-rich galaxies along the
sightlines to 48 nearby Lyman-$\alpha$ absorbers (collaborators M. Putman, E.
Ryan-Weber, J. Stocke). Here we will restrict our discussion to the Arecibo 
HI observations around 16 low redshift ($cz < 12750$ \kms) absorbers which span
a range in column density 12.81 $< log$ N$_{HI} < 14.09$ cm$^{-2}$
\citep{penton2004, penton2000} assuming b = 25 \kms. 

The Arecibo observations cover a 31.2\arcm $\times$ 31.2\arcm field around each
of the 4 QSO sightlines that were observed (PG 1211+143, PG 1116+215, Mrk 335, and
Ton 1542). The rms noise in the data cubes after Hanning smoothing is $\sim$
0.0015 -- 0.002 Jy/beam. In these fields we detect 8 new gas-rich galaxies (we 
call any galaxy with no previously known redshift a ``new" system even if they 
were previously cataloged) with HI masses ranging from 1.4$\times 10^8$ M\solar\ to
4.6$\times 10^9$ M\solar\ and we re-detect 3 large spiral galaxies (IC 3061, NGC
4529, and CGCG 126-027).

We do not detect any galaxies in the immediate vicinity of the absorption
sightlines. The nearest HI-rich galaxy is 95 kpc from the absorber implying
that, at least to the sensitivity limits of this survey, the halos of gas-rich 
dwarf and LSB galaxies are not responsible for Lyman-$\alpha$ forest absorption 
unless their gaseous halos are extremely large relative to their luminosity. 
Figure 3 shows the 
large scale galaxy distributions around each of these QSO sightlines. The 
Xs mark the positions of the Lyman-$\alpha$ absorbers, the triangles mark 
the positions of a previously known galaxies that were re-detected, the large
filled circles mark the positions of the newly detected gas-rich galaxies, and 
the smaller filled circles mark the positions of galaxies identified in ZCAT
(http://cfa-www.harvard.edu/~huchra/zcat/). This figure shows that most of the
Lyman-$\alpha$ forest absorbers fall along the large scale structures defined the
the optical and 21 cm galaxy distributions. However, while 8 of the 16 absorbers 
are within 500 kpc of a galaxy, there are 2 Lyman-$\alpha$ absorbers in the
sample that are more than 2 Mpc from the nearest identified galaxy clearly
putting them in voids. 

\begin{figure}[ht]
\begin{center}
\leavevmode
\epsfxsize=2.6in
\epsffile{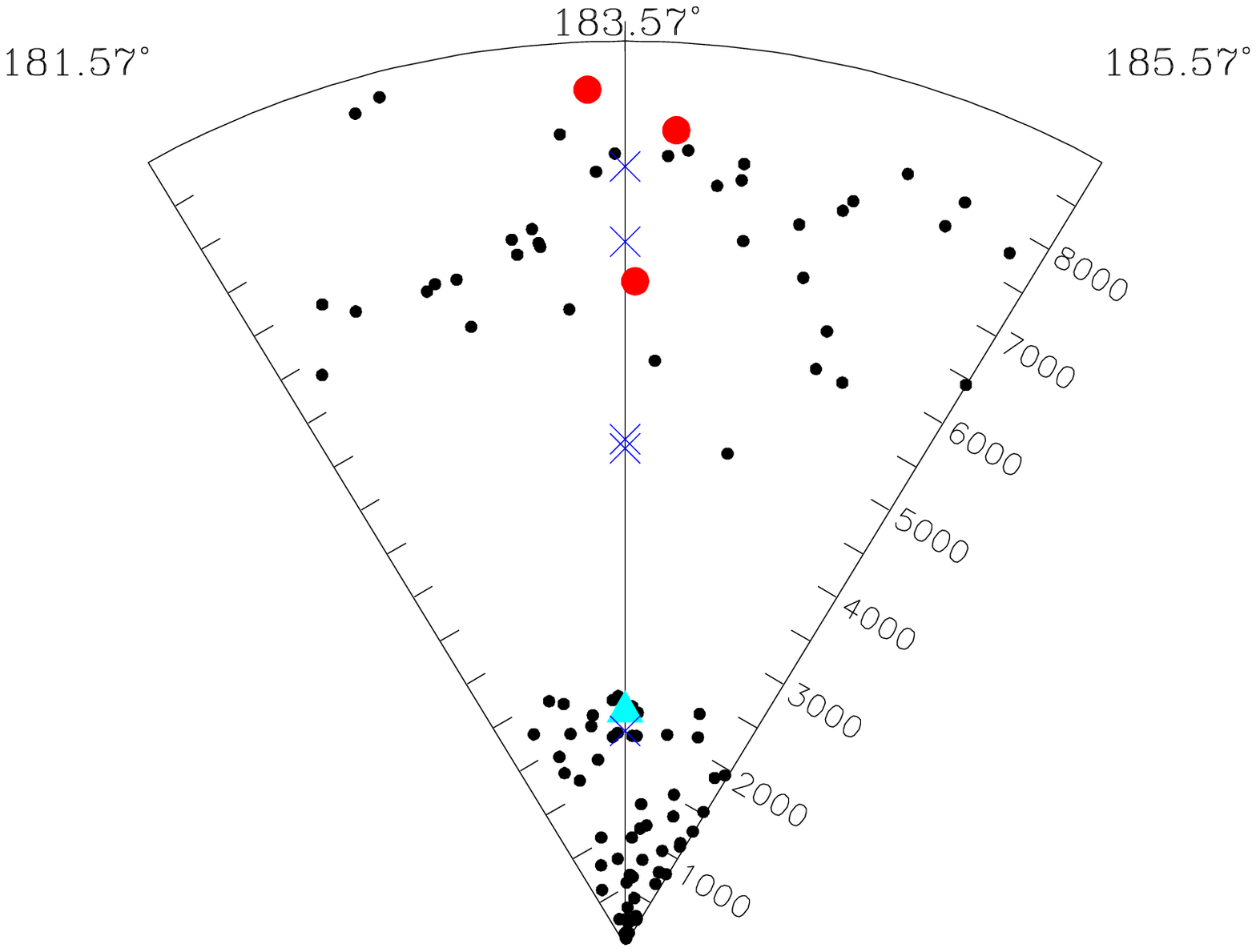}
\epsfxsize=2.6in
\epsffile{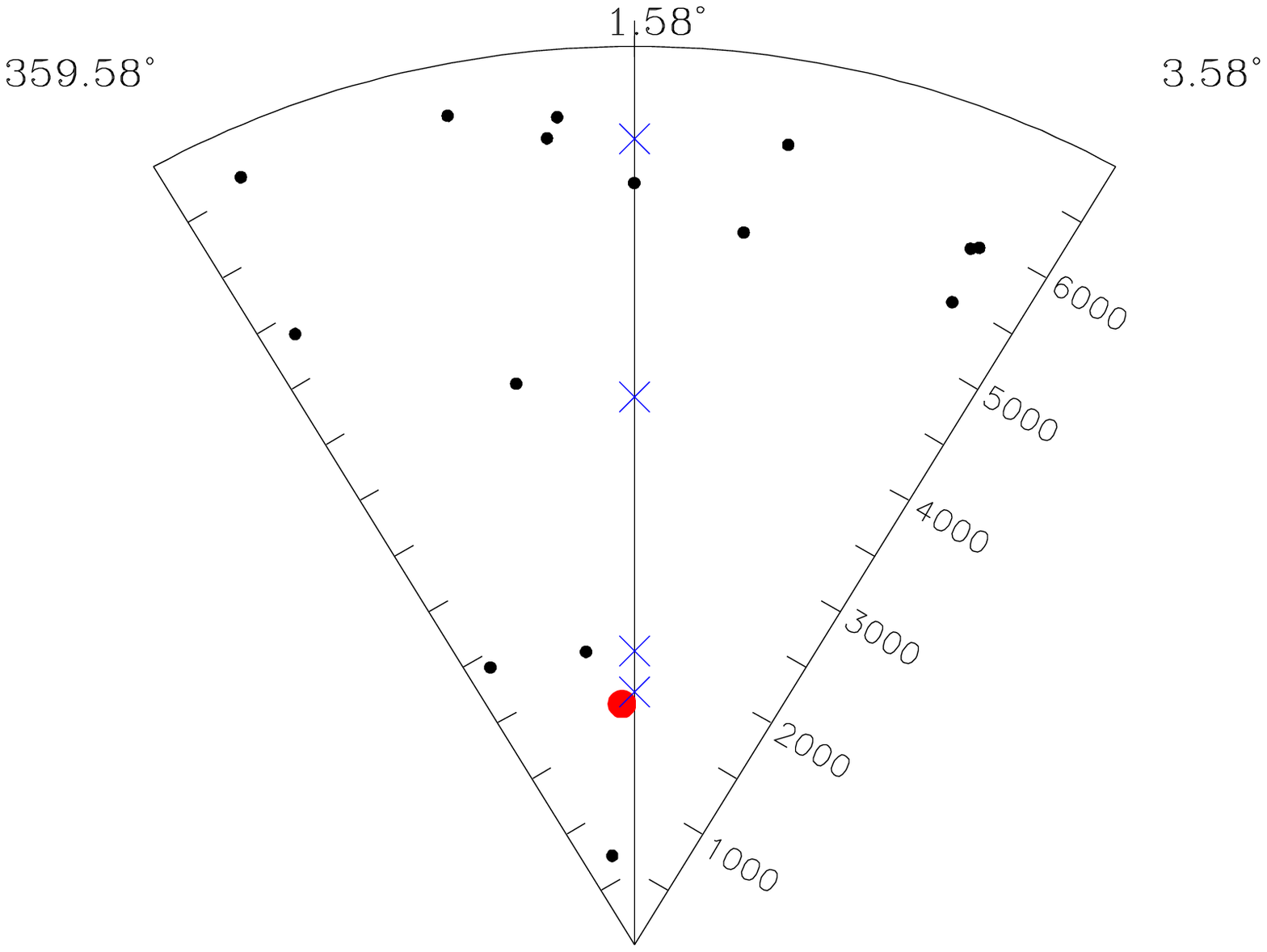}
\end{center}
\begin{center}
\leavevmode
\epsfxsize=2.6in
\epsffile{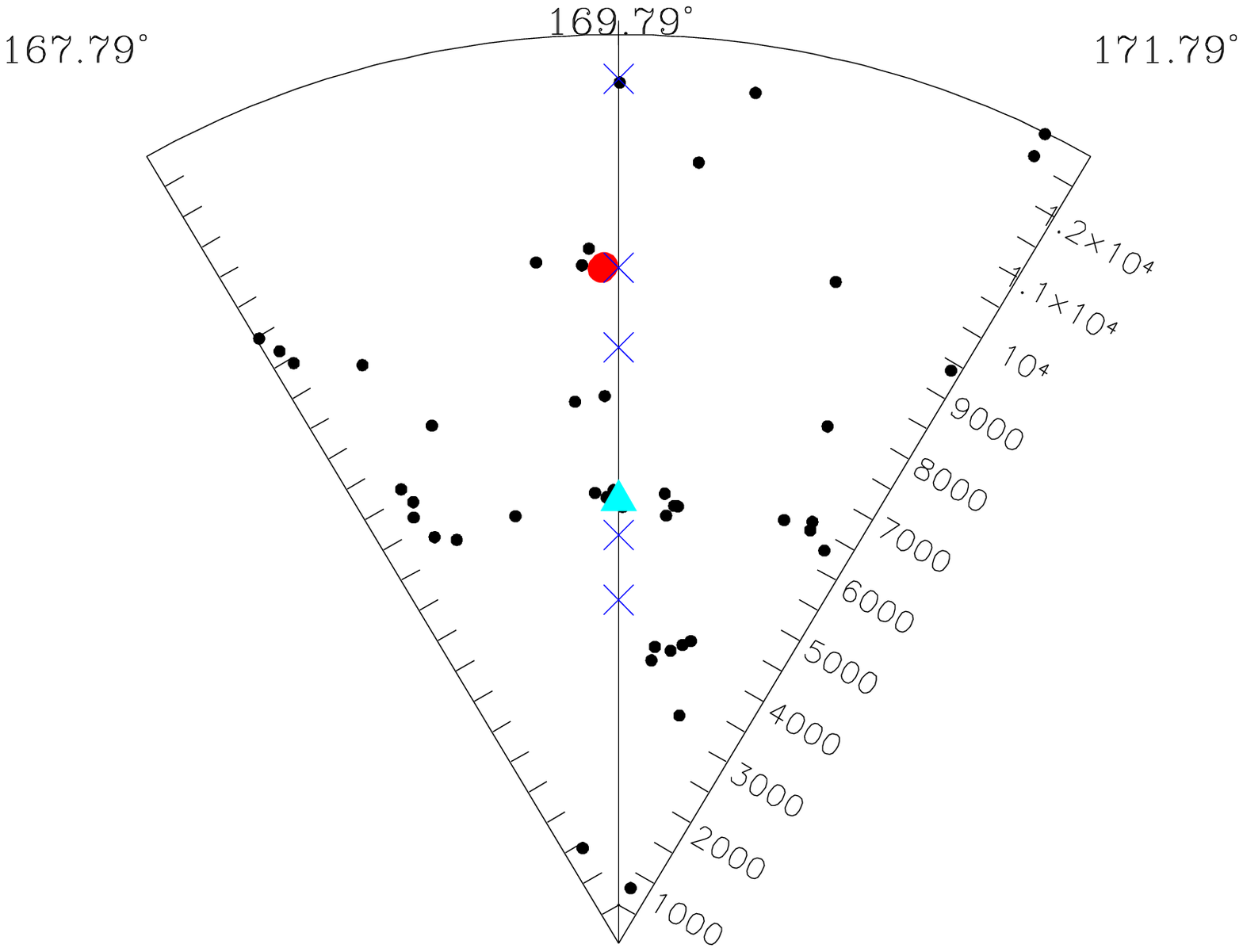}
\epsfxsize=2.6in
\epsffile{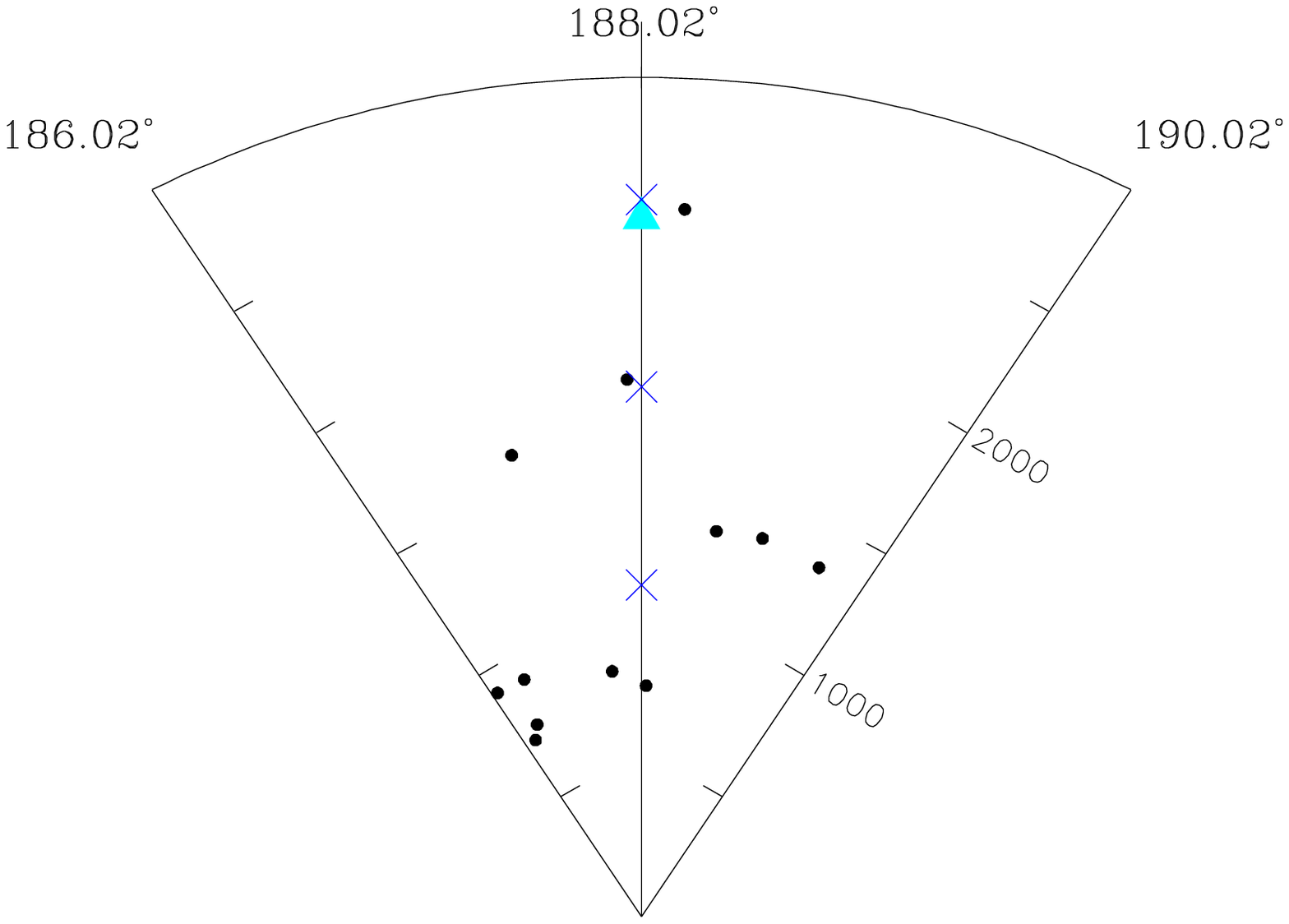}
\end{center}
\caption{The large scale distribution of galaxies around PG1211 and Mrk 335
(top) and PG1116 and Ton 1542 (bottom). The
Xs mark the positions of the Lyman-$\alpha$ absorbers along these sightlines,
the triangles are the locations of bright spiral galaxies that we detect at 21
cm but were previously known. The large circles are new detections of gas-rich
galaxies in this survey. Some of them were previously known, but none had
previously cataloged redshifts. The small filled circles are the galaxies 
listed in the ZCAT survey consisting of the CfA redshift data and other galaxy 
catalogs that have been included (http://cfa-www.harvard.edu/~huchra/zcat/).}
\end{figure}

\section{Summary and Conclusions}

HI observations allow for the detection of faint, LSB
companions to Lyman-$\alpha$ absorbers of all sorts. These observations provide
an excellent probe of the small scale HI environment of these absorbers that is
only traceable in the local universe. As shown in these studies, there are
many gas-rich dwarf and LSB galaxies in the local universe
that have gone undetected in optical spectroscopic surveys. 

The nearest DLA has been studied in detail \citep{bowen2001a, bowen2001b,
chengalur2002}, yet it was not previously known that it resides in a very
gas-rich environment with several close companions. Nevertheless this appears 
to be a very young region resembling those that should be more prevalent in the 
higher redshift universe. This region shows that without the
ability to detect these low luminosity, gas-rich systems at higher redshift we
may be missing significant clues to understanding the nature of the regions in
which DLAs reside.

Despite a deep search for galaxies along the lines of sight to
16 Lyman-$\alpha$ absorbers, we only detected 11 gas-rich galaxies all of which
were $>$ 95 kpc from the absorber. While many of these absorbers follow the
large scale structure traced by galaxies in the region, there are some that
appear to reside in voids at least 2 Mpc from the nearest known galaxy. Given
the lack of detection of any galaxy within 95 kpc of these absorbers it appears
that the absorption does not arise in the halos of dwarf or low surface 
brightness galaxies missed in optical surveys.

\acknowledgments

The work presented here is a compilation of a couple of projects. The
collaborators on these projects have made important contributions to this work
and to the results published here. For their contributions I
acknowledge the work of David Bowen, Todd Tripp, Mary Putman, Emma Ryan-Weber, 
and John Stocke. I also acknowledge the financial support for this work provided
by the NSF Astronomy and Astrophysics Postdoctoral Fellowship program under grant 
AST-0302049.


\end{document}